\documentclass[twocolumn, trackchanges]{aastex631}

\shortauthors{Diop et al.}

\graphicspath{{./}}

\begin{document}
\title{Disentangling CO Chemistry in a Protoplanetary Disk using Explanatory Machine Learning Techniques}

\author{Amina Diop}
\affiliation{Department of Astronomy, University of Virginia, Charlottesville, VA 22904, USA}

\author{L. Ilsedore Cleeves}
\affiliation{Department of Astronomy, University of Virginia, Charlottesville, VA 22904, USA}

\author{Dana E. Anderson}
\affiliation{Earth and Planets Laboratory, Carnegie Institution for Science, 5241 Broad Branch Road, NW, Washington, DC 20015, USA}
\altaffiliation{Carnegie Postdoctoral Fellow}

\author{Jamila Pegues}
\affiliation{Space Telescope Science Institute, Baltimore, MD 21218, USA}
\altaffiliation{STScI Postdoctoral Fellow}

\author{Adele Plunkett}
\affiliation{National Radio Astronomy Observatory, 520 Edgemont Rd., Charlottesville, VA 22903 USA}

\begin{abstract}
Molecular abundances in protoplanetary disks are highly sensitive to the local physical conditions, including gas temperature, gas density, radiation field, and dust properties. Often multiple factors are intertwined, impacting the abundances of both simple and complex species. We present a new approach to understanding these chemical and physical interdependencies using machine learning. Specifically we explore the case of CO modeled under the conditions of a generic disk and build an explanatory regression model to study the dependence of CO spatial density on the gas density, gas temperature, cosmic ray ionization rate, X-ray ionization rate, and UV flux. Our findings indicate that combinations of parameters play a surprisingly powerful role in regulating CO compared to any singular physical parameter. Moreover, in general, we find the conditions in the disk are destructive toward CO. CO depletion is further enhanced in an increased cosmic ray environment and in disks with higher initial C/O ratios. These dependencies uncovered by our new approach are consistent with previous studies, which are more modeling intensive and computationally expensive. Our work thus shows that machine learning can be a powerful tool not only for creating efficient predictive models, but also for enabling a deeper understanding of complex chemical processes.
\end{abstract}

\section{Introduction} \label{sec:intro}
The chemistry of protoplanetary disks, like the chemistry of the interstellar medium (ISM), is highly sensitive to the local physical conditions. However, the densities, high radiation fields, and dust growth in disks lead to unique chemical pathways that are not reflected in the ISM. To interpret line observations and effectively use molecules as tracers of important properties such as the disk mass or the cosmic ray ionization rate, it is crucial to determine how the wide range of physical conditions in disks impact chemical abundances.

CO, for example, has been recently shown to be more sensitive to the disk environment than initially thought. Specifically, gas-phase CO is found to be sub-interstellar in disks when comparing CO-derived gas masses with HD and dust-derived masses \citep{bergin13,favre13,ansdell16,mcclure16,Long17,miotello17, bergwil18,calahan21} even after accounting for freeze-out in the cold midplane, UV-driven photodissociation in the surface layers, and isotope-selective photodissociation \citep{miotello14,miotello17}. This  suggests that there are additional chemical processes impacting CO. Models suggest that CO can be chemically processed by cosmic rays and X-rays \citep[e.g.,][]{reboussin15,Bosman18,dodson18, schwarz18,schwarz19}. These studies also find that CO processing is influenced by the temperature structure, UV radiation field, and dust distribution, confirming the significant impact of the disk environment on the chemistry.

A common approach to study how CO and other molecular species change with the environment is to generate a grid of full 2D chemical models with varying disk and stellar properties \citep[e.g.,][]{walsh10,walsh12,reboussin15,schwarz18,schwarz19}. This approach has greatly informed our understanding of molecules and has been especially helpful when interpreting observations. However, generating such a grid quickly becomes computationally expensive, which makes it challenging to carry out a detailed exploration of parameter space.

In the present paper, we explore whether complementing chemical modeling efforts with machine learning (hereafter ML) techniques can enable us to explore large parameter spaces and analyze large model data sets more efficiently and unveil relationships that deepen our understanding of chemical processes in disks. We focus on CO since, over the years, we have gained a solid understanding of how its chemistry is impacted by  the environmental conditions, which provides a point of comparison for  our purely ML-driven approach. ML is a subset of artificial intelligence that uses data-driven algorithms to unveil trends in large amounts of data. In recent years, ML has gained popularity in the field of astronomy given the increasing complexity and amount of astronomical data (see Baron \citeyear{baron19}). It has been used, for example, to classify stellar spectra \citep{sanchez13}, detect unusual galaxies \citep{baron17}, derive star formation rates \citep{veneri19}, calculate masses of protoplanets \citep{auddy20,shangjia2022}, detect protoplanets in disks \citep{Terry2022}, search for exoplanets \citep{exominer22}, and analyze the spectral energy distributions of protoplanetary disks \citep{kauefer23}.

Our goal is to assess the ability of ML in helping us understand the interplay between chemistry and physical conditions using CO as a test case. We focus on the {\em explanatory} power of ML rather than its predictive power, i.e., we aim to build a reliable predictive model that can also help us understand the relationship between the inputs (disk physical conditions) and the output (CO abundance) and whether these inputs favor the production or destruction of CO. The predictive power of ML has fully been exploited in recent studies to speed up astrochemical modeling \citep[e.g.,][]{mijolla19,holdship21, grassi21, smirnov22}. For our ML data set, instead of generating a grid of 2D chemical models, we focus on a single model since disks inherently have within them a diversity of environments due to the temperature, density, and radiation gradients.

\section{Methods}\label{sec:methods}
\subsection{Disk modeling} \label{sec:chem_mod}
We adopt a chemical model published in \citet{anderson21}. This model is not intended to represent the diversity of known disks (e.g., MAPS I \citep{maps1}), but rather, it is a generic model that highlights the range of physical conditions naturally present within a single disk, which allows us to sample a large parameter space and investigate its impact on CO chemistry.  Essentially, we aim to leverage the diversity of chemical environments within the disk to reveal how the CO abundance (the output of the ML model) and the physical conditions (the inputs of the ML model) are linked with a regression-based analysis. We summarize here the main physical and chemical features of the model, but we direct the reader to \citet{anderson21} for further details.

The disk structure consists of a  two-dimensional azimuthally symmetric disk with a gas mass of 0.01 M$_\sun$ and a dust mass of 10$^{-4}$ M$_\sun$. The disk surrounds a 1 M$_\sun$ T Tauri star with a radius of 2.8 R$_\sun$, an effective temperature of 4300 K, and an X-ray luminosity of 10$^{30}$ erg s$^{-1}$. We use two dust populations: a small dust population (0.005 - 1 $\mu$m) and a large dust population (0.005 $\mu$m - 1 mm),  which are made up of 80 $\%$ astronomical silicates and 20 $\%$ graphite \citep{draine84}. Each dust population follows an MRN size distribution \citep{mathis77}. The large dust population contains 90 $\%$ of the total dust mass, while  the remaining mass is made up of the small dust population. The radial surface densities for the gas and dust are respectively described by:
\begin{equation}
    \Sigma_{\textrm{g}}(R) = 23\; \textrm{g}\; \textrm{cm}^{-2}  \biggl(\frac{R}{R_c}\biggr)^{-1}\textrm{exp}\biggl[-\frac{R}{R_c}\biggr]
\end{equation}

\begin{equation}
    \Sigma_{\textrm{d}}(R) = 0.23\; \textrm{g}\; \textrm{cm}^{-2} \biggl(\frac{R}{R_c}\biggr)^{-1}\textrm{exp}\biggl[-\frac{R}{R_c}\biggr]
\end{equation}
where R$_c$ = 30 au. The scale height for the gas and small grains is given by:
\begin{equation}
   \textrm{H}(R) = 10\;\textrm{AU} \biggl(\frac{R}{100 \; \textrm{AU}} \biggr)^{1.15}
\end{equation}
The scale height of the large grains is five times lower in order to simulate dust settling, following \citet{andrews11}. The densities for the small and large dust populations are given by:
\begin{equation}
    \rho_s(R,Z) = \frac{(1-f)\Sigma_d}{\sqrt{2\pi}H}\textrm{exp}\biggl[-\frac{1}{2}\biggl(\frac{Z}{H} \biggr)^2\biggr]
\end{equation}

\begin{equation}
    \rho_l(R,Z) = \frac{f\Sigma_d}{\sqrt{2\pi}\chi H}\textrm{exp}\biggl[-\frac{1}{2}\biggl(\frac{Z}{\chi H}\biggr)^2\biggr]
\end{equation}
where f = 0.9 represents the fraction of mass in the large grains and $\chi$ = 0.2 denotes the ratio between the scale height of the large grains to that of the small grains. \

Radiative transfer is performed using TORUS \citep{Harries00, Harries04, Kurosawa04, Pinte09} to calculate dust temperatures given the assumed gas and dust structure. The attenuation for the UV and X-ray radiation fields is carried out using the Monte Carlo radiative transfer code by \citet{Bethell11a,bethell11b}. The gas temperatures are estimated based on the local UV flux and gas density using fitting functions to thermochemical models of \citet{bruderer13} as outlined in \citet{cleeves15}. For the cosmic ray ionization, we use the solar system minimum model from \citet{cleeves15}.  The incident cosmic ray ionization rate is 1.1 $\times$ 10$^{-18}$ s$^{-1}$ and is modulated with vertical depth \citep[see][]{cleeves15}.

\begin{figure*}[htb!]
\gridline{\fig{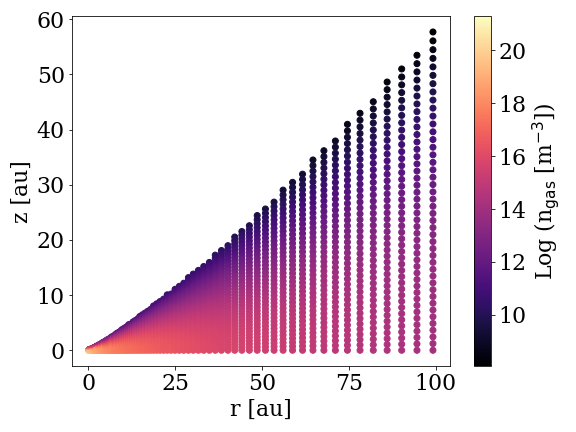}{0.33\textwidth}{(a)}
\fig{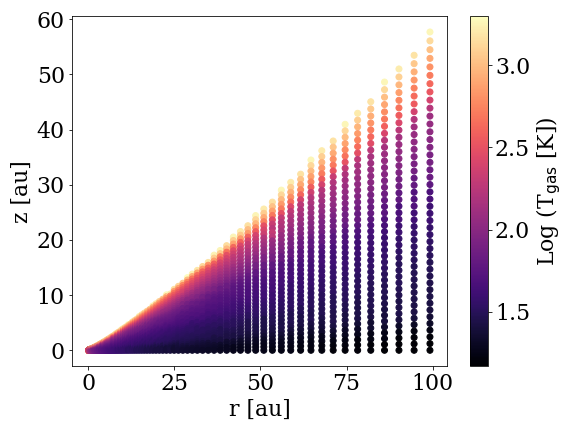}{0.33\textwidth}{(b)}
\fig{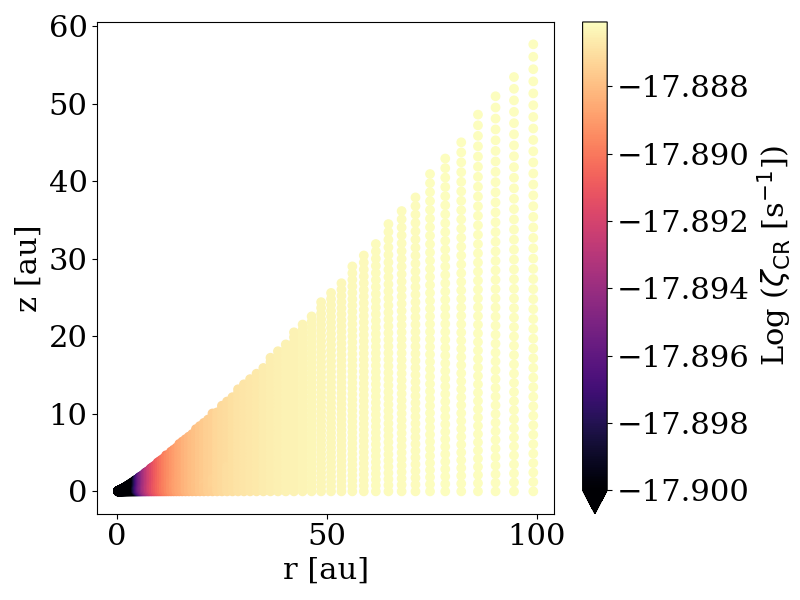}{0.33\textwidth}{(c)}}
\gridline{\fig{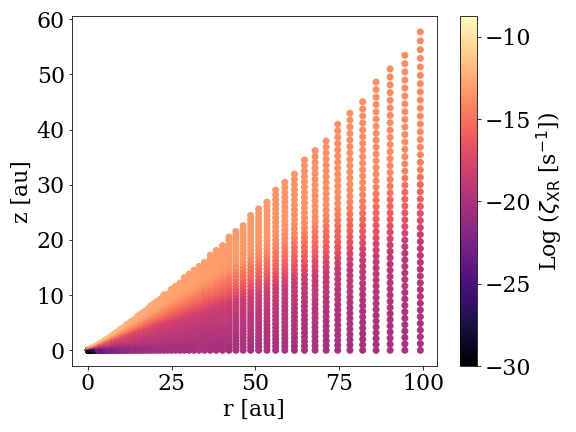}{0.33\textwidth}{(d)}
\fig{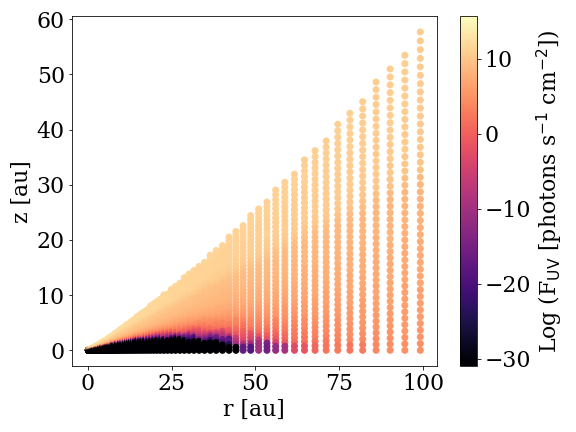}{0.33\textwidth}{(e)}
\fig{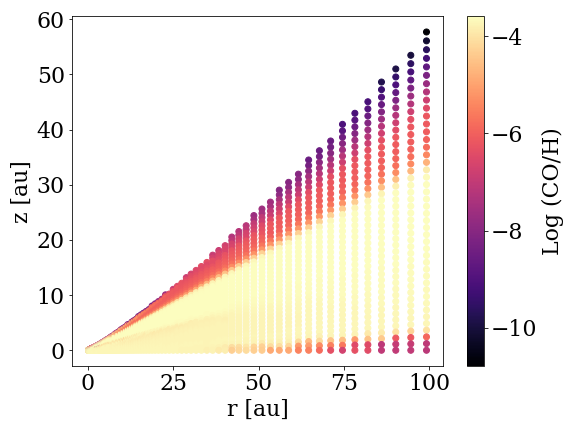}{0.33\textwidth}{(f)}}
\caption{Physical structure and CO abundance as a function of (r, z) for the fiducial disk model. The disk extends radially from 0.2 au to 100 au and vertically from 0 to 60 au. The CO abundance evolves over time, while the physical structure remains static.  All the quantities are shown on a logarithmic scale (base 10). (a) Gas density (b) Gas temperature (c) Cosmic ray ionization rate. Note that the disk is largely transparent to cosmic rays, except within the inner few au. (d) X-ray ionization rate (e) UV flux (f) CO abundances at $\sim$ 1 Myr obtained from the chemical calculations. The abundances are expressed relative to the total amount of hydrogen, i.e. H + 2H$_2$. Most locations have abundances around 10$^{-4}$ per H atom.}
\label{fig:phys_param_co}
\end{figure*}

The model is sampled on a grid of 130 radii between 0.2 and 100 au, logarithmically spaced, and 50 heights linearly spaced in z/r (0 $<$ z/r $<$ 0.6) to capture the wide diversity of chemical environments spanning the disk. Figure  \ref{fig:phys_param_co} illustrates how the physical parameters vary from location to location in the disk model.

The abundance of CO as a function of time is calculated with the time-evolving gas-grain chemical code of \citet{Fogel11}, which is based on the Ohio State University gas-phase network \citep{smith04}. The code has been subsequently updated in \citet{cleeves13,cleeves14,cleeves15, cleeves18} and \citet{anderson21}. The abundances are computed by applying the rate equation method. Our network is comprised of 654 species and 7039 reactions and includes neutral-neutral reactions, ion-neutral reactions, ion recombination with electrons, a limited number of grain surface reactions, freeze-out, cosmic ray ionization,  thermal and non-thermal desorption, and photodissociation including self-shielding for H$_2$, CO, and N$_2$ \citep{Lee96, Visser09, Li13}. The abundances are calculated for 100 time steps between 1 yr and 3 Myr. The initial abundances are shown in Table \ref{tab:abund}. The disk has initial C/H and O/H ratios of  1.3 $\times$ 10$^{-4}$ and  2.28 $\times$ 10$^{-4}$ respectively, which corresponds to an initial C/O ratio of 0.57. Figure \ref{fig:phys_param_co} (panel f) shows the results of the chemical calculations for CO at 1 Myr.

In addition to the fiducial disk model described above, we explore two more chemical models, (1) a model with an ISM-like cosmic ray ionization rate of $\sim$ 2 $\times$ 10$^{-17}$ s$^{-1}$ \citep{w98} and (2) a model with a higher initial C/O ratio (0.83) obtained by lowering the initial H$_2$O abundance to 8 $\times 10^{-6}$. The latter model simulates water-rich ice sequestration motivated by \citet{hogerheijde11} and relatively high inferred C/O ratios in observed disk systems \citep{bergin16,cleeves18,Miotello19}.

\begin{deluxetable}{cccc}
\tablecaption{Initial abundances (relative to hydrogen) for the disk chemical models.Ices are indicated by the notation\emph{gr}.
\label{tab:abund}}
\tablehead{\colhead{Molecule} & \colhead{Abundance} & \colhead{Molecule} & \colhead{Abundance}}
\decimals
\startdata
H$_2$ &  5.00 $\times 10^{-1}$ & He & 1.40 $\times 10^{-1}$ \\
CO & 9.92 $\times 10^{-5}$  &  H$_2$O (gr) & 8.00 $\times 10^{-5}$ \tablenotemark{*}\\
N$_2$ & 3.51 $\times 10^{-5}$ & CO$_2$ (gr) & 2.24 $\times 10^{-5}$\\
NH$_3$ (gr) & 4.8 $\times 10^{-6}$ & CH$_3$OH (gr) & 4.8 $\times 10^{-6}$ \\
CH$_4$ (gr)& 3.6 $\times 10^{-6}$ & HCN & 2.00  $\times 10^{-8} $ \\
H$_3^+$ & 1.00 $\times 10^{-8}$  &  HCO$^{+}$ & 9.00 $\times 10^{-9}$\\
C$_2$H & 8.00 $\times 10^{-9}$ & SO & 5.00 $\times 10^{-9}$ \\
CS& 4.00 $\times 10^{-9}$  & Mg$^+$ & 1.00 $\times 10^{-11}$ \\
Fe$^+$ & 1.00 $\times 10^{-11}$ &  Si$^+$  & 1.00 $\times 10^{-11}$\\
\enddata
\tablenotetext{*}{The high C/O model has an abundance of 8 $\times 10^{-6}$.}
\end{deluxetable}

\subsection{Machine learning approach} \label{sec:ML}
\begin{figure*}[htb!]
    \centering
    \includegraphics[width = \textwidth]{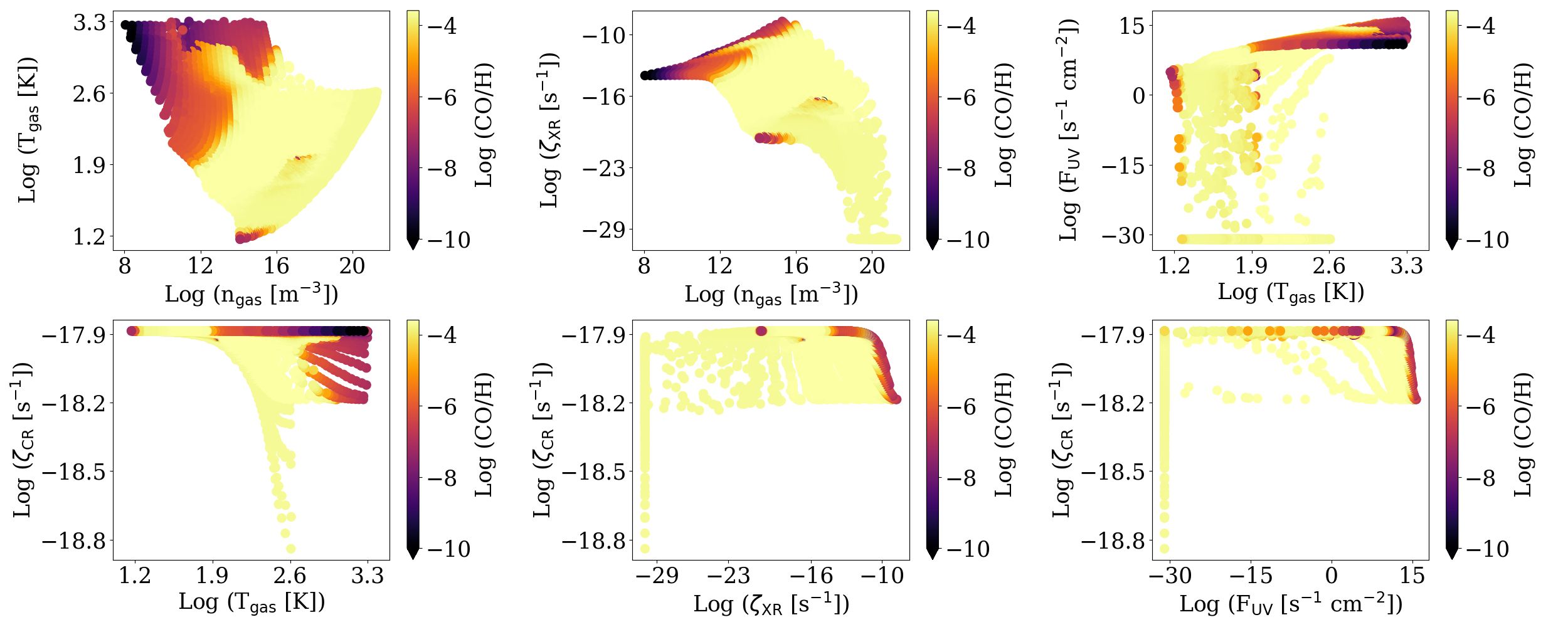}
    \caption{CO abundance at $\sim$ 1 Myr as a function of different pairs of predictors. We note a smooth variation across parameter space for each plot, yet the relationship between CO and the entire set of predictors is complex.}
    \label{fig:co_dens_pair}
\end{figure*}

We aim to understand the complex web of dependencies between the physical environment of the disk and the resulting CO chemistry. Instead of taking a full 2D disk model as a single sample, we treat every (r,z) point location  as a sample of the CO environment (Figure \ref{fig:phys_param_co}). As mentioned previously, CO abundance depends on many variables, however Figure \ref{fig:co_dens_pair} illustrates that while complex, the CO abundances vary smoothly across parameter space, allowing us to effectively apply ML to the CO problem.

\subsubsection{ML data}\label{ml_data}
 The data for the regression model is made up of the grid of points representing the full disk (Figure \ref{fig:phys_param_co}). Each point has its own gas temperature (T$_{\textrm{gas}}$), gas density (n$_{\textrm{gas}}$), X-ray ionization rate ($\zeta_{\textrm{xr}}$), cosmic ray ionization rate ($\zeta_{\textrm{cr}}$), and UV flux (F$_{\textrm{uv}}$), which represent the independent variables, often called predictors or features in the field of ML. The CO abundances relative to hydrogen (CO/H) extracted at $\sim$ 1 Myr also vary across points and represent the dependent variable.

 \subsubsection{Preparing the data}
\begin{figure*}[htb!]
    \centering
    \includegraphics [width=\textwidth]{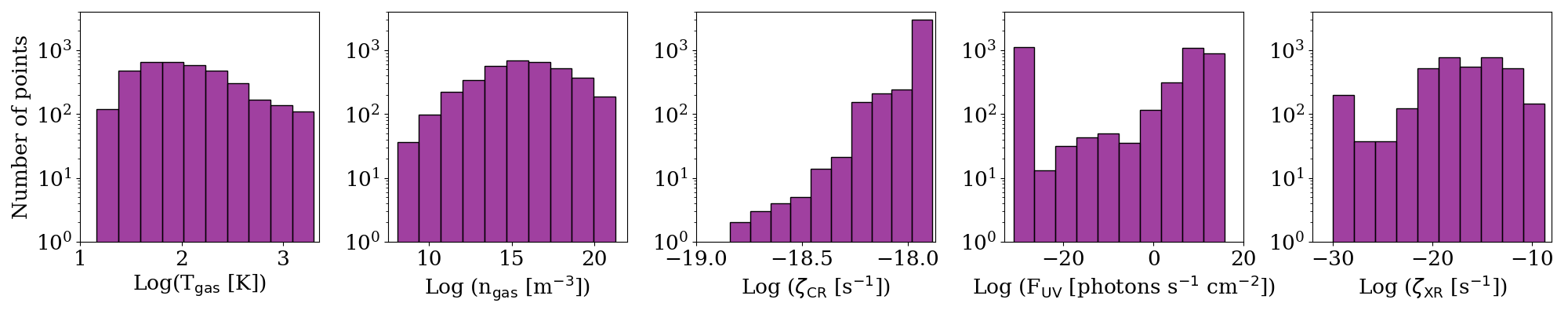}
    \caption{Parameter space sampling for the independent variables: gas temperature, gas density, cosmic ray ionization rate, UV flux, and X-ray ionization rate. Besides the cosmic ray ionization rate, the parameter space samples multiple orders of magnitude in the values of these physical quantities.}
    \label{fig:dist}
\end{figure*}

We build the regression model using the Python libraries Scikit-learn \citep{scikit} designed for ML tasks and Statsmodels \citep{statsmodels}, which is used for exploratory statistical analysis. During data cleaning, we excluded points where n$_{\rm{gas}}$ or n$_{\rm{CO}}$ = 0. After cleaning, the final number of data points is 3688 and their distribution in parameter space is shown in Figure~\ref{fig:dist}.   Since the data span many orders of magnitude in both independent and dependent variable values, we work in a base 10 logarithmic space. We randomly split the data set into a training set and a test set using the Scikit-learn function train$\_$test$\_$split. The training set is fed into the ML algorithm so that it can learn the trend. The test set simulates data that the trained model has never encountered and is used to test the model performance.

We follow a 70$/$30 split: we use 70 $\%$ of the data for training and 30 $\%$ for testing. After splitting, we perform a feature scaling on the predictors/independent variables to ensure the coefficients of the regression model have the same units. This scaling step allows us to directly determine the significance of the predictors compared to each other. We use the robust scaling technique, defined as:
\begin{equation}
    x_i' = \frac{x_i -\textrm{median}(x)}{Q_{3}(x) -Q_1(x)}
\end{equation}
where x$_i$ is the value of the predictor for the i$^{\text{th}}$ point,  x$'_i$ the scaled value, median(x) the median value of the predictor, and $Q_{3}(x) - Q_{1} (x)$ the interquartile range. This technique is robust against outliers and does not assume a normal distribution for the predictors. To prevent data leakage, which occurs when the test set information is available during the training phase, we exclusively use the training set to compute scaling parameters (median and interquartile range), which are subsequently used to scale the test set.

\subsubsection{ML algorithms: multiple linear regression and polynomial regression}
In this work, we use multiple linear regression and polynomial regression. A multiple linear regression is a model where the dependent variable Y and the predictors X$_i$= $\{X_1, X_2,...,X_n\}$ are related via:

\begin{equation}
    Y = \beta_0 + \beta_1 X_1 + \beta_2 X_2 + ...+ \beta_n X_n + \epsilon
\end{equation}
where $\beta_0$ is the intercept, $\beta_1$, $\beta_2$, ... $\beta_n$ the coefficients, and $\epsilon$ the error term. A polynomial regression has a similar form, except that it contains higher order terms and can contain cross-terms:
\begin{equation}
    Y = \beta_0 + \beta_1X_1 + \beta_{11} X_1^2 + \beta_2X_2  + \beta_{12} X_1X_2  + ... + \epsilon
\end{equation}
For instance, in this work, Y corresponds to log$_{10}$(CO abundance), while X$_1$ and X$_2$ correspond to the scaled log$_{10}$(gas density) and log$_{10}$(gas temperature). We feed the scaled training data to the ML algorithm, which searches for the best regression model, i.e., the model with the set of $\beta$ values that minimize the error. We use the coefficients of the resulting model to determine the impact of the disk physical conditions on CO.

\subsubsection{Evaluating model performance }\label{sec:eval}
\begin{figure*}[htb!]
    \centering
    \includegraphics[width=\textwidth]{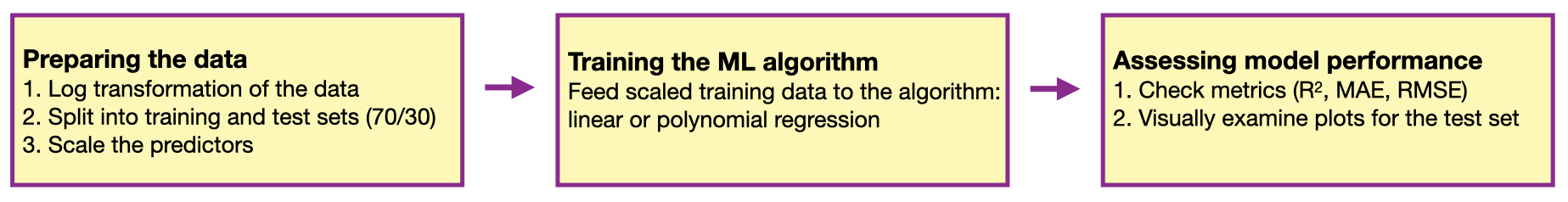}
    \caption{Flowchart summarizing our ML approach. We first apply a base 10 logarithmic transformation to the independent variables and the dependent variable since they span many orders of magnitude. The data are then split into a training set and a test set. The independent variables/predictors are then scaled to ensure that the coefficients of the regression model have the same units. The training data is fed to the ML algorithm, in this case, linear or polynomial regression. The performance of the resulting model is assessed using the metrics described in Section~\ref{sec:eval}. We also visually examine residual plots and plots of the predicted (ML model) vs actual (chemical model) CO abundances after applying the trained ML algorithm to the test set.}
    \label{fig:ml_workflow}
\end{figure*}

We assess the performance of the model using the following metrics: the coefficient of determination or R$^2$ score, the root mean square error (RMSE), and the mean absolute error (MAE) defined as,
\begin{equation}
    \text{R}^2 =1- \frac{\sum_i(y_i -\hat{y}_i)^2}{\sum_i(y_i -\bar{y})^2}
\end{equation}

\begin{equation}
   \text{RMSE} = \sqrt{\frac{\sum_{i=1}^N(y_i -\hat{y}_i)^2}{N}}
\end{equation}

\begin{equation}
   \text{MAE} = \frac{1}{N}\sum_{i=1}^N|y_i -\hat{y}_i|
\end{equation}
where $y_i$ is the actual CO abundance (from the chemical model), $\hat{y_i}$ the abundance predicted by the regression model, $\bar{y}$ the mean of the actual abundances, and \emph{N} the number of data points. The R$^2$ score ranges between 0 and 1 and describes the proportion of variation in the dependent variable that can be predicted from the independent variables. The closer the R$^2$ is to 1, the better the model. The RMSE and MAE are used to gauge the error of the model. To further evaluate the performance of the model, we visually examine residual plots and plots of the predicted vs actual CO abundances using the test set. Figure \ref{fig:ml_workflow} summarizes our ML approach.

\section{Results} \label{subsec:res}
\subsection{Molecular density is more robustly predicted than molecular abundance}\label{abundvsdens}
\begin{figure*}[htb!]
    \centering
    \includegraphics[width = \textwidth]{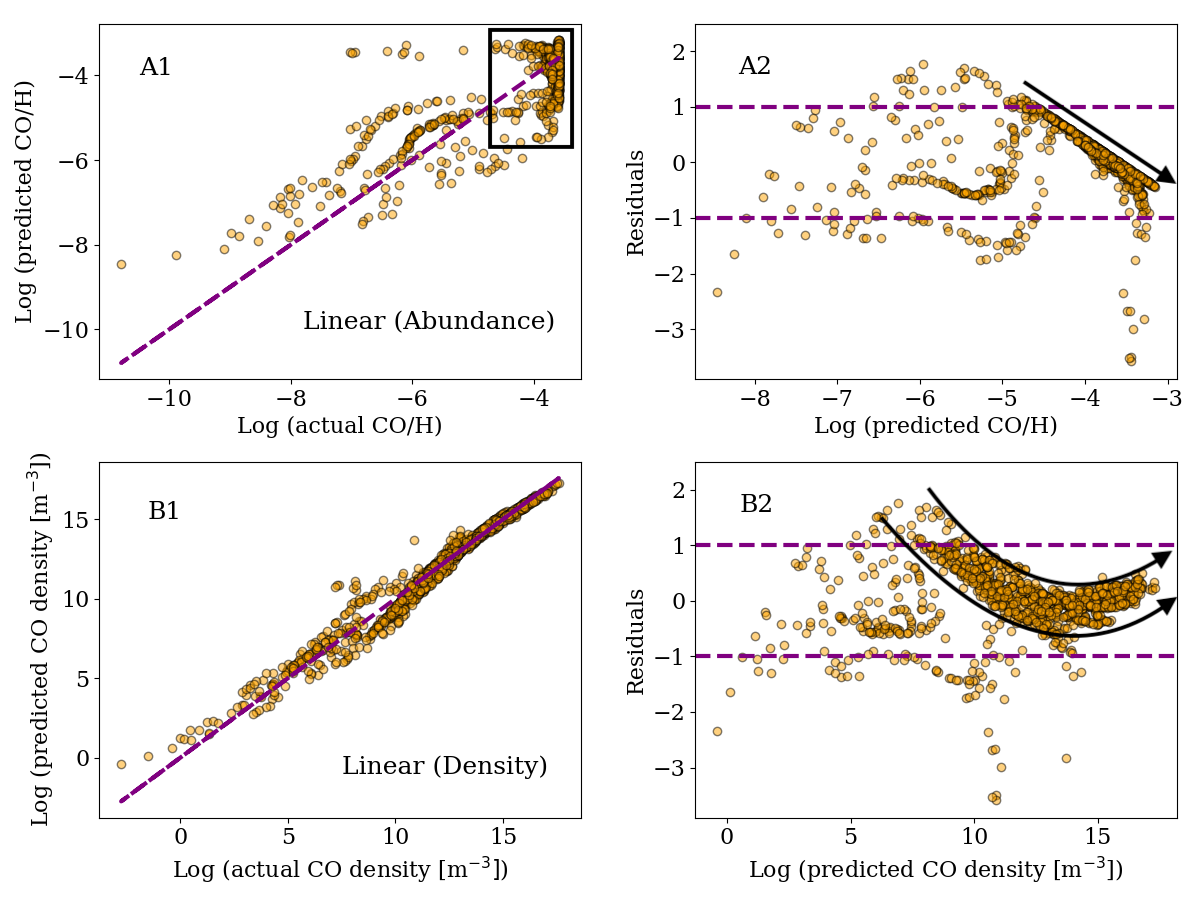}
    \caption{(\textbf{A}) Results of the multiple linear regression in abundance space using the test set. A1: Predicted (ML model) vs actual (chemical model) CO abundances on a log-log plot. The purple line indicates the one-to-one correspondence. The black rectangle shows the region where there is a cluster of points due to the cap on the CO abundance in the chemical model. A2: Residuals vs predicted CO abundances. The residuals are defined as the difference between log (actual CO abundance) and log (predicted CO abundance). The black arrow shows the linear trend due to the cluster. The dashed purple lines indicate a difference of 1 dex between the actual and predicted abundance. (\textbf{B}) Same as panel (A) except CO densities are used. The residuals are defined as the difference between log (actual CO density) and log (predicted CO density). Using densities removes the cluster, which significantly improves the quality of the linear fit. However, some curvilinear trends (indicated by black arrows) are noted in the residual plot (B2), thus a more complex model is needed.}
    \label{fig:abund_dens_reg}
\end{figure*}

For the multiple linear regression, we initially use T$_{\textrm{gas}}$, n$_{\textrm{gas}}$, $\zeta_{\textrm{xr}}$, $\zeta_{\textrm{cr}}$, and F$_{\textrm{uv}}$ (see Section \ref{ml_data}) as our predictors and CO abundance (CO/H) as our dependent variable. We find that working with the abundance as the output challenges the ability of the algorithm to find a good linear model. This behavior is due to the fact that the chemical model naturally has a maximum value for the CO abundance at the limits of the C and/or O atomic budget and many CO-rich points have abundances close to this value. This aspect of the chemical model leads to the edge cluster in Figure \ref{fig:abund_dens_reg} (panel A1). To address this issue, we work instead in the smoother CO density space, i.e., abundance $\times$ hydrogen density,  which is continuous and naturally  removes the cluster as shown in Figure \ref{fig:abund_dens_reg} (panels B1 and B2). The shift from CO abundance to CO density results in an improvement in the R$^2$ score, from 0.72 to 0.97.  We note that this change causes CO density to be strongly predicted by the gas density, which must be taken into account in the interpretation of the coefficients of the ML model (see Section \ref{coeff_analysis}).

\subsection{Even ``simple'' CO chemistry requires  polynomial regression} \label{pol_reg}

While a linear model is generally preferred due to its simplicity, we find that it is not sufficiently accurate for predicting CO densities. Specifically, as shown in Figure \ref{fig:abund_dens_reg} (panel B2), the residual plot reveals curvilinear trends, which suggests that the linear model does not fully capture the underlying CO behavior.

\begin{figure*}[htb!]
    \centering
    \includegraphics[width = \textwidth]{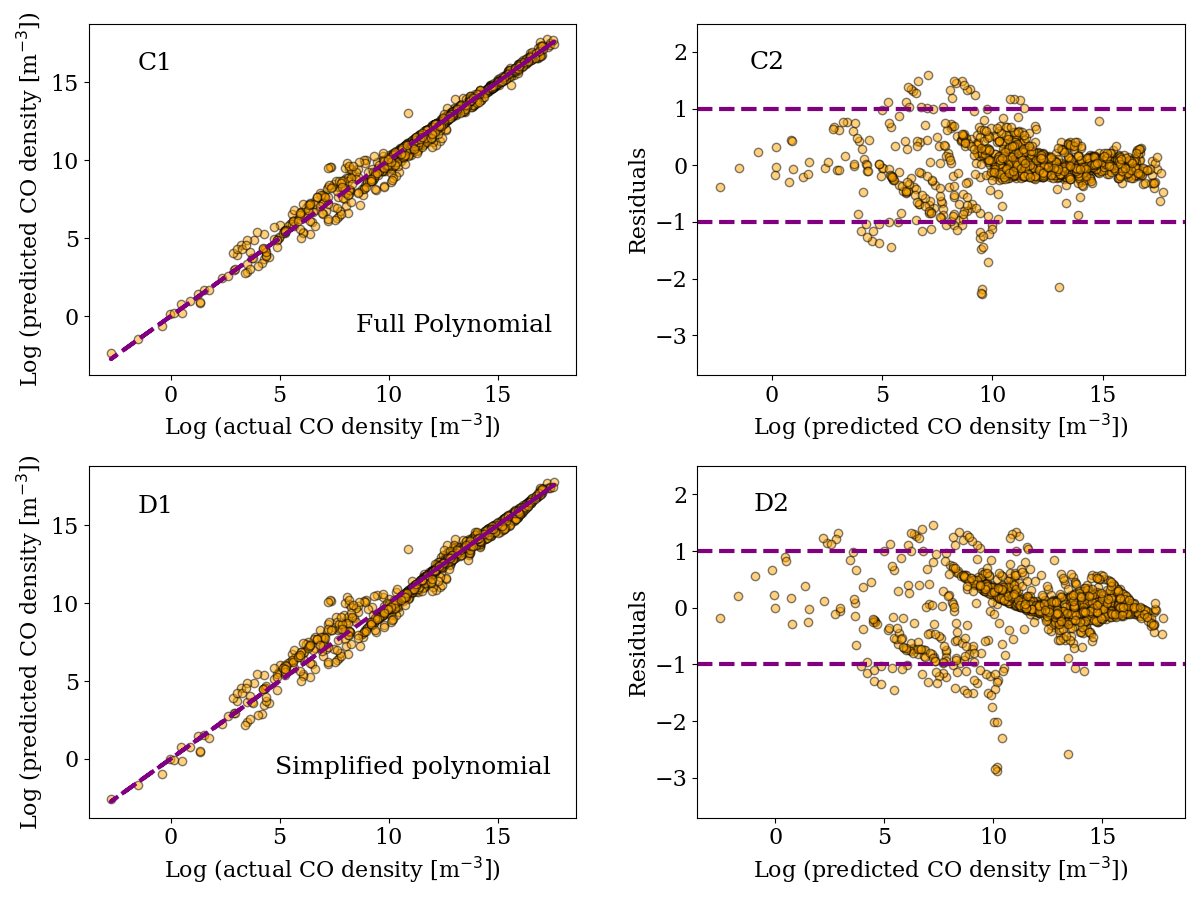}
    \caption{(\textbf{C}) Results of the polynomial regression using the test set. The full polynomial model fits the data better than the linear model and curvilinear trends in the residuals are reduced. (\textbf{D}) Results of the simplified polynomial. The full polynomial matches better the data, but the simplified polynomial facilitates interpretation and does not suffer as much from multicollinearity since it contains fewer terms.}
    \label{fig:full_simp_pol}
\end{figure*}

Instead, we find that a second degree polynomial provides a substantial improvement in our ability to produce accurate CO densities. Figure \ref{fig:full_simp_pol} (panels C1 and C2) shows the results of the polynomial model generated using the function PolynomialFeatures in Scikit-Learn. This model shows a good agreement between the predicted and actual CO densities and curvilinear trends are reduced. The predicted densities, especially between 10$^{-2.5}$ and 10$^{10}$ m$^{-3}$, match more closely the actual values in the polynomial model. The residual plot also shows that for most points, the predicted densities are off by 0.5 dex and the maximal deviation is $\sim$ 2 dex. Quantitative metrics likewise reveal that the polynomial is a better model: Table \ref{tab:metrics_lin_full_simp} shows that it has a slightly higher R$^2$ value as well as smaller RMSE and MAE values than the linear model. Thus, the second-degree polynomial provides a more reliable description of CO chemistry in the disk model.

\subsection{Simplified polynomial regression: how many terms are necessary?}\label{how_many_terms}
The aim of this work is to create both an accurate and interpretable model that unveils the dependence of CO chemistry on the physical environment. To this end, we seek to analyze the coefficients of the terms in the ML model to determine the different processes affecting CO and whether these processes lead to its destruction or preservation. The polynomial model does capture the behavior of CO under the modeled disk conditions, but it is comprised of 20 terms, which makes interpretation challenging. Moreover, the terms suffer from multicollinearity, i.e., the independent variables/predictors are correlated with one another. Multicollinearity poses a challenge for our goal of building an explanatory model since it makes it difficult to isolate the impact of each predictor on CO without the influence of the remaining predictors. We thus reduce the number of terms in the polynomial to minimize the impact of multicollinearity and obtain a simplified model that is easier to interpret. To do this, we use a combination of variation inflation factor analysis, centering, and feature importance scores from random forests (see Appendices \ref{rfs} and \ref{mult}).

The results of the simplified polynomial model are shown in Figure \ref{fig:full_simp_pol} (panels D1 and D2). While the full polynomial model does have better a performance as shown by the plots (Figure \ref{fig:full_simp_pol}), its slightly higher R$^2$ score and lower RMSE and MAE values (Table \ref{tab:metrics_lin_full_simp}), the simplified model contains fewer terms, captures most of the underlying behavior of CO, and does not suffer as much from multicollinearity. Thus, we focus on this model for the remainder of the study.

\begin{figure}[t]
    \centering
    \includegraphics[scale=0.33]{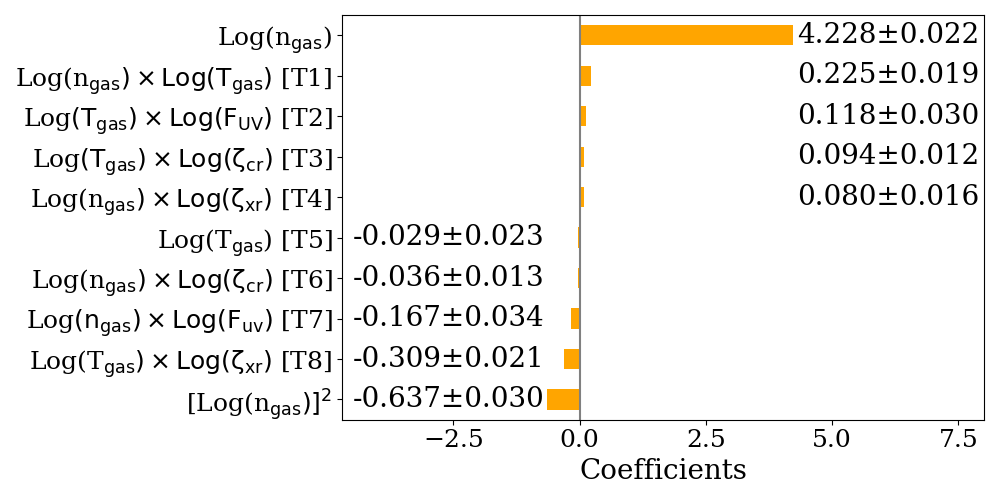}
    \caption{Coefficients of the simplified polynomial with uncertainties calculated by statsmodels.}
    \label{fig:simplified_coeff2}
\end{figure}

The simplified model is made up of 10 terms with their coefficients shown in Figure \ref{fig:simplified_coeff2}. The dominant terms are log(n$_{\rm{gas}}$) and [log(n$_{\rm{gas}})]^2$, which are partly influenced by the use of CO density (the product of CO/H and n$_{\rm{gas}}$) instead of CO abundance. The remaining terms involve combinations of factors with T$_{\rm{gas}}$ and n$_{\rm{gas}}$ being the factors most commonly found in the cross-terms.

\begin{deluxetable*}{cccccc}
\tablecaption{Metrics for the linear, full polynomial, and simplified polynomial models of the full disk and the linear model of the warm molecular layer. The full polynomial model is a better fit than the linear model given that it has a slightly higher R$^2$ score and smaller RMSE and MAE values. The full polynomial is a better model than the simplified polynomial since it has a slightly higher R$^2$ score and lower RMSE and MAE values. However, the latter does not suffer as much from multicollinearity and is easier to interpret. \label{tab:metrics_lin_full_simp}}
\tablehead{\colhead{Model} & \colhead{R$^2$ score} & \colhead{RMSE (Train)} & \colhead{RMSE (Test)} & \colhead{MAE (Train)} & \colhead{MAE (Test)}}
\decimals
\startdata
Linear & 0.971 & 0.599 &  0.593 & 0.393 & 0.402\\
Full polynomial & 0.986 & 0.418 &  0.410 & 0.249 & 0.258\\
Simplified polynomial & 0.98 & 0.496 &  0.490 & 0.313 & 0.323\\
Linear (warm region) & 0.970 & 0.313 &  0.273 & 0.118 & 0.120\\
\enddata
\end{deluxetable*}

\section{Discussion}\label{subsec:disc}
\subsection{Understanding CO chemistry using the coefficients of the ML model}\label{coeff_analysis}
Our simplified polynomial model encodes a substantial amount of information about the factors affecting CO. Its terms reveal what processes are at play and whether they destroy (negative terms) or preserve CO (positive terms). Here, we unpack the information contained in each term.
\subsubsection{Dominant terms: $\log (n_{\rm {gas}})$ and $[\log (n_{\rm gas})]^2$} \label{gas_dens}
The strongest correlation our analysis identifies is between CO and the gas density terms, $\log (n_{\rm {gas}})$ and $[ \log (n_{\rm gas} ) ]^2$. This correlation is naturally expected since CO scales with H$_2$ in disks \citep{bergwil18,obergberging21}. Our use of CO density as the dependent variable further strengthens this correlation. The negative relationship between CO density and $[ \log (n_{\rm gas} ) ]^2$ is at first glance surprising, but this term reflects the fact that denser regions tend to be colder and have higher degrees of UV extinction, resulting in CO freeze-out into ice.

\subsubsection{Positive correlations}
The terms with positive coefficients hint at processes that lead to the preservation of CO in the gas phase. The model contains four positive terms that we will label T1, T2, T3, and T4 as follows: $\log (n_{\rm {gas}}) \times \log(T_{\rm{gas}})$ (T1), $\log(T_{\rm{gas}}) \times  \log(F_{\rm{UV}})$ (T2), $\log(T_{\rm{gas}}) \times \log(\zeta_{\textrm{cr}})$ (T3), and  $\log(n_{\rm{gas}}) \times  \log(\zeta_{\rm{xr}})$ (T4). T1 is picking up regions where $n_{\rm{gas}}$ is sufficiently high to shield CO from UV radiation and $T_{\rm{gas}}$ is high enough to prevent CO freeze-out. T2 corresponds to warm regions with moderate UV fields, which help maintain CO in the gas-phase. Since the cosmic ray rates do not vary substantially throughout the disk (Figure \ref{fig:phys_param_co}), T3 could reflect (1) warm regions where CO is in the gas phase and cosmic rays produce sufficient He$^+$ to keep CO from forming more complex species or (2) cool ($\lesssim20$~K) midplane regions where cosmic rays induce non-thermal desorption \citep{hasegawa93}.
Meanwhile, the positive correlation between CO and T4 is intriguing since high X-ray rates and densities should facilitate the formation of radicals that destroy CO. This term could be reflecting a zone with a  sufficiently high X-ray ionization rate that destroys complex species that form out of CO, thus helping bring back CO to its original state.

\subsubsection{Negative correlations}
The negative terms $\log (T_{\rm {gas}})$ (T5), $\log (n_{\rm {gas}}) \times  \log(\zeta_{\textrm{cr}})$ (T6), $\log (n_{\rm {gas}}) \times  \log(F_{\rm{UV}})$ (T7), and $\log (T_{\rm{gas}}) \times \log(\zeta_{\textrm{xr}})$ (T8) highlight pathways that lead to CO destruction. T5 reveals a negative correlation between CO density and temperature, which could be highlighting the hot surface layers (T$_{\rm{gas}}$ $>$ 100 K) where CO is photodissociated. T6 captures regions where high densities and cosmic rays destroy CO: (1) regions where cosmic ray ionization leads to He$^+$, which destroys CO gas and ultimately leads to the formation of hydrocarbons such as CH$_4$ (2) denser midplane regions where cosmic rays produce atomic hydrogen from the ionization of H$_2$, which enables the hydrogenation of CO in the ice-phase, ultimately forming species like CH$_3$OH.
T7 is picking up a region of higher density, which causes more frequent gas-grain collisions. With a moderate UV field present, radicals like OH and CH$_3$ can be efficiently formed, which can react with more volatile species like CO and convert them over time into more complex molecules that freeze out, such as CH$_3$OH or CO$_2$ \citep{schwarz18}.
Finally, T8 is picking up warm, irradiated regions which are relatively harsh environments where X-rays can be another source of He$^+$.

\subsubsection{What do we learn from the coefficients?}
The coefficient analysis reveals that CO correlates strongly with the amount of gas, but it is impacted by multiple processes that can either enhance it or destroy it. Disregarding the gas density terms $\log(n_{\rm{gas}})$ and [$\log(n_{\rm{gas}})]^2$ for the reasons previously discussed, we  note that the remaining negative terms have coefficients that outweigh the positive ones, suggesting a tendency toward CO depletion in the disk model. Our analysis also points to the importance of considering combinations of physical conditions, which can trigger or hinder different chemical processes, in sometimes counterintuitive ways. {\em Considering combinations of physical parameters} is especially important when conducting sensitivity analyses of molecular species. In addition, the cross-terms typically involve T$_{\rm{gas}}$ and n$_{\rm{gas}}$, which confirms previous results suggesting that the gas temperature and gas density structures are crucial factors regulating CO chemistry.

\subsection{What about the warm molecular layer?}
\begin{figure*}[htb!]
    \centering
    \includegraphics[width =\textwidth]{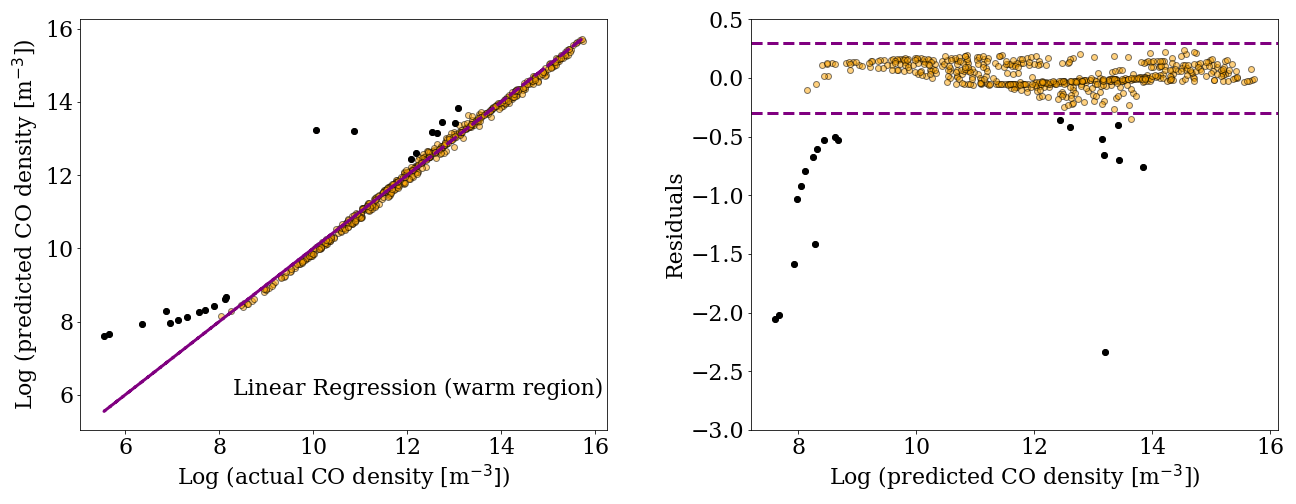}
    \caption{Results of the multiple linear regression on the warm molecular layer (20 K $<$ T$_{gas}$ $<$ 100 K ) using the test set. The outliers (21/538 points) with residuals $>$ 3 $\sigma$ ($\sigma$ = MAE) are colored in black. Most points lie along the line of one-to-one correspondence and have residuals between -0.3 and 0.3 dex as indicated by the dashed purple lines in the residual plot. This indicates that the linear model captures CO chemistry in the warm molecular layer.}
    \label{fig:reg_warm}
\end{figure*}

Our analysis up to this point focuses on the entire disk model, but CO observations usually probe the warm molecular layer where the temperatures maintain CO in the gas-phase and are expected to reduce depletion. We investigate this by subsampling the locations (1792/3688 points) associated with the warm molecular layer (20 K $<$ T$_{\rm{gas}}$ $<$ 100 K) and look for a model that captures CO chemistry in this region. We perform  a multiple linear regression with T$_{\textrm{gas}}$, n$_{\textrm{gas}}$, $\zeta_{\textrm{xr}}$, $\zeta_{\textrm{cr}}$, and F$_{\textrm{uv}}$ as the predictors and note a strong correlation between the predicted and actual densities (Figure \ref{fig:reg_warm}). There are a few outliers (21/538 samples in the test set) with residuals $>$ 3 $\sigma$ ($\sigma$ = MAE), which explains the asymmetric residual map. Nevertheless, most of the residuals lie between -0.3 and 0.3 dex.

\begin{figure*}
    \centering
    \includegraphics[width = \textwidth]{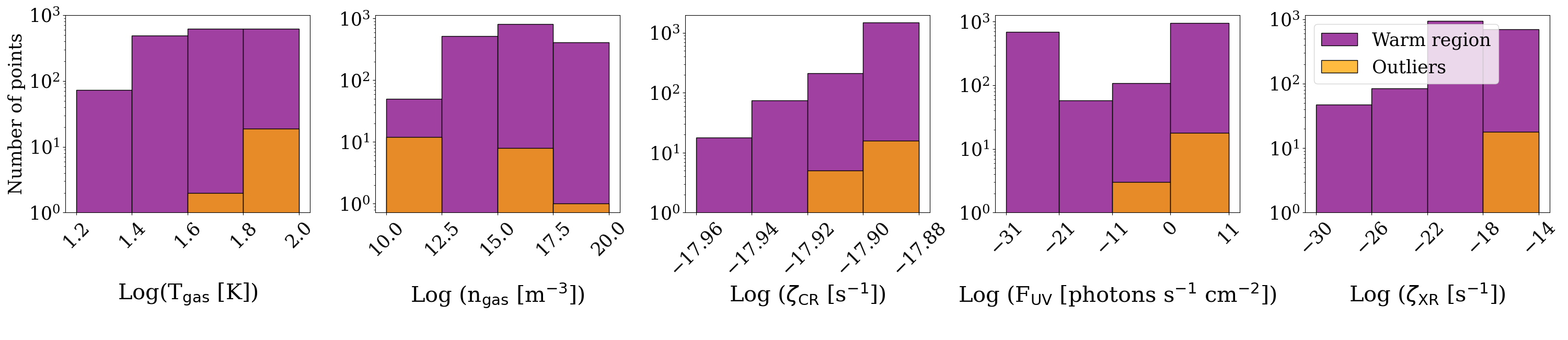}
    \caption{Parameter space sampling for the warm molecular layer (purple) and the outliers in that region (orange; see Figure \ref{fig:reg_warm}). The outliers occupy regions close to the edges of parameter space where the model is less well sampled. They coincide with highly photo-irradiated gas, where CO is undergoing a rapid transition via UV photodissociation  or water is sublimating rapidly (near 100 K), and thus there are not many points sampling these regions.}
    \label{fig:warm_outliers}
\end{figure*}

Within the small number of outliers, most of them lie at the edges of parameter space (Figure \ref{fig:warm_outliers}) and are located in the transition region between the atomic and molecular layers in the r-z plane. Some outliers also occupy hot regions, $\sim$ 100 K, where water begins to desorb, changing the oxygen chemistry in the gas and impacting CO. Both regions, i.e., the transition region where photodissociation starts to become important and the hot region with gas-phase water,  undergo rapid chemical changes, where not many points exist to sample their behavior. Thus, it is not surprising the algorithm struggles to reproduce CO chemistry at these locations without additional data.

On the whole, the multiple linear regression model does well capturing CO chemistry in the warm molecular layer. This suggests that CO is more robust in this sub-region as compared to the full extent of the disk, which requires a polynomial model. Warmer temperatures (20 K $<$ T$_{\rm{gas}}$ $<$ 100 K) thus seem to reduce the complexity of CO chemistry by hindering its chemical destruction. The implication of this finding is that predictive models of CO chemistry should be more accurate for warm regions (e.g., for disks orbiting more massive Herbig stars), compared to cooler regions including the disks of lower mass stars.

\subsection{Additional model variations: cosmic ray ionization rate, initial C/O ratio,  and age}
\begin{figure*}[htb!]
    \centering
    \includegraphics[width=0.95\textwidth]{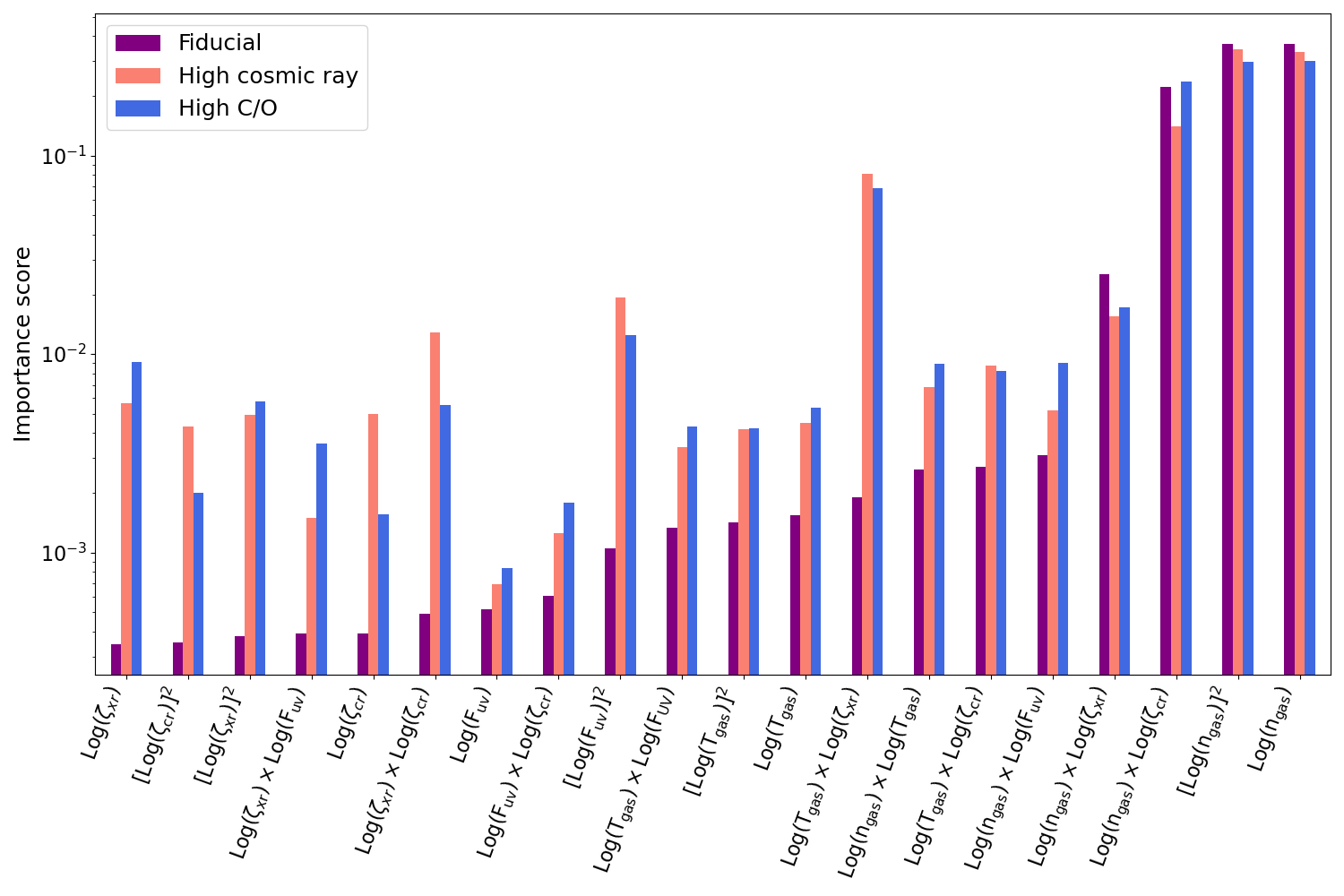}
    \caption{Feature importance scores for the fiducial model, high cosmic ray model ($\sim$ 2 $\times 10^{-17}$ s$^{-1}$), and high C/O model (0.83) at $\sim$ 1 Myr. The three models have the same dominant terms (log(n$_{\rm{gas}}$), [log(n$_{\rm{gas}})]^2$, and log(n$_{\rm{gas}}$) $\times$  log($\zeta_{\rm{cr}}$)), but the cross-terms involving the radiation field and the cosmic rays are more important in the high cosmic ray and high C/O models.}
    \label{fig:feat_model_comp}
\end{figure*}

We explore additional chemical models to investigate the impact of using an ISM-like cosmic ray ionization rate ($\sim$ $10^{-17}$ s$^{-1}$), an increased initial C/O ratio (0.83 instead of 0.57), and different ages compared to 1 Myr ($\sim$ 0.5, $\sim$ 1.5, and $\sim$ 2 Myr). We use the feature importance  score from random forests  to determine the importance of the terms in each model (see Appendix \ref{rfs} for details). For all of these variations, the relative behaviors of the dominant terms do not change substantially. For all three models, i.e., the fiducial, high cosmic-ray, and high C/O models at 1 Myr, the dominant terms are the log(n$_{\rm{gas}}$), [log(n$_{\rm{gas}})]^2$, and log(n$_{\rm{gas}}$) $\times$  log($\zeta_{\rm{cr}}$) terms (Figure \ref{fig:feat_model_comp}). This consistent behavior implies that predicting CO chemistry is fairly robust even as disk physical conditions vary.

\subsubsection{An ISM-like cosmic ray rate enhances CO depletion}
For the high cosmic ray model, the cross-terms involving the radiation field and cosmic rays are more important than in the fiducial model. This can be explained by the fact that the higher rate leads to more H$_3^+$ and He$^+$ ions, which destroy CO to ultimately form species such as CH$_3$OH and CH$_4$ respectively \citep{schwarz18,schwarz19}.

\subsubsection{A higher initial C/O enhances CO depletion}
For the high C/O model, the same cross-terms as in the high cosmic ray model are more important than in the fiducial model. This effect could be due to the fact that the lower initial H$_2$O abundance creates an oxygen-poor environment, which increases the chances of some of the carbon being locked up in hydrocarbons such as C$_2$H instead of CO, which further reduces CO's ability to self-shield \citep{bergin16, Miotello19}. With the reduced shielding, more CO can be dissociated by the radiation field, thus freeing more carbon atoms, which are needed for C$_2$H formation.

\subsubsection{CO depletion increases over time}
\begin{figure*}[htb!]
    \centering
\includegraphics[width=0.95\textwidth]{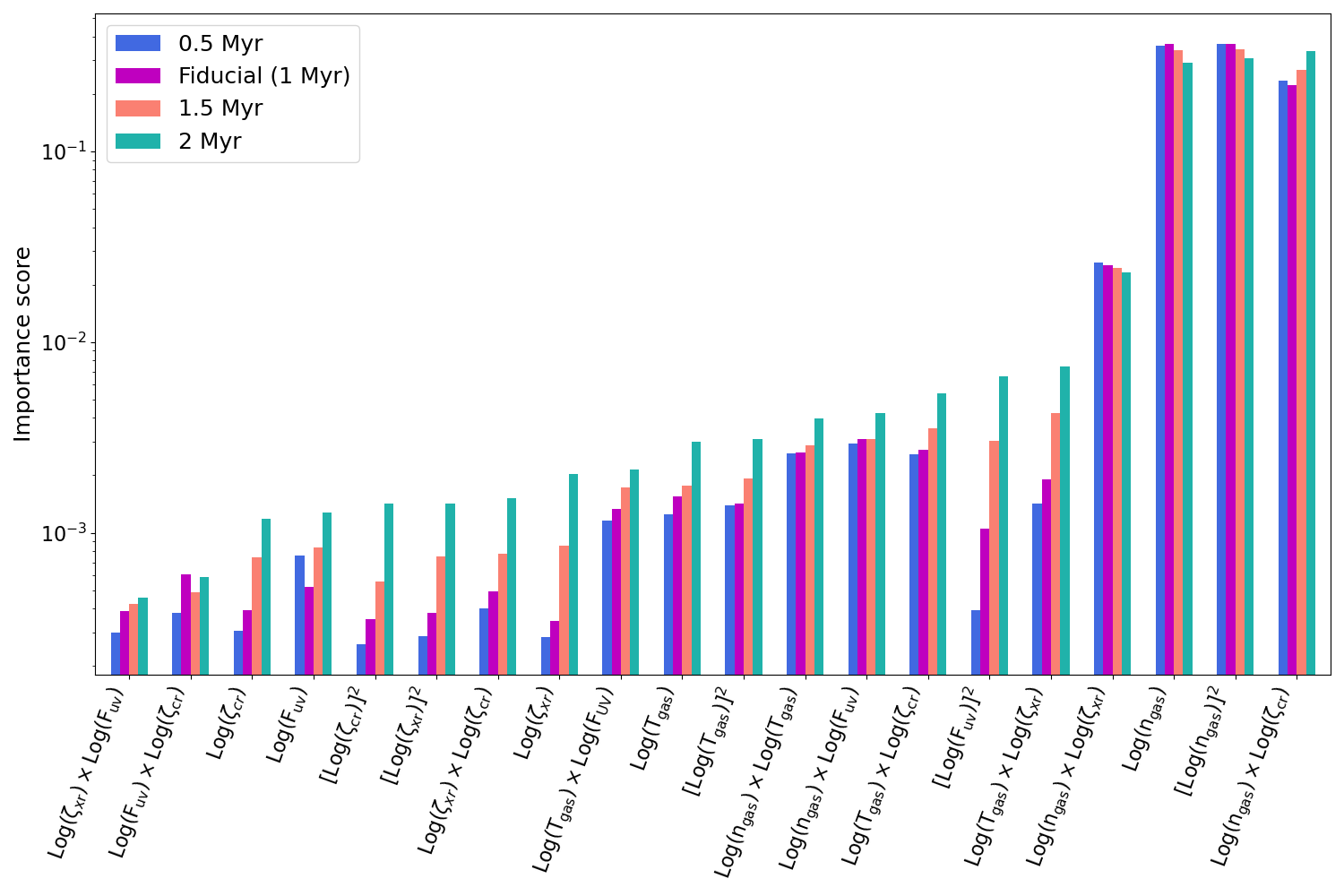}
    \caption{Feature importance scores for the fiducial model at different ages. Over time, the term log(n$_{\rm{gas}}$) $\times$  log($\zeta_{\rm{cr}}$) starts to dominate and cross-terms involving the radiation field and cosmic rays become more important.}
    \label{fig:fid_time}
\end{figure*}

Over the 2 Myr timescale, the term log(n$_{\rm{gas}}$) $\times$  log($\zeta_{\rm{cr}}$) becomes the most important term in the disk model (Figures \ref{fig:fid_time}) and the cross-terms involving cosmic rays and the radiation field become more important. This can be explained by the timescales needed to convert CO into species such as CO$_2$, CH$_4$, CH$_3$OH \citep{reboussin15}. Over time, the abundance of these species is enhanced because they will have had enough time to form thanks to the processing of CO, which usually involves the radiation field and cosmic rays.

\subsection{Putting it all together: performance of ML approach}
Our purely ML-driven approach leads to findings consistent with what we have learned about CO chemistry over the years,  notably, (1) our fiducial model shows a tendency toward CO depletion, which is in agreement with the CO depletion trend commonly observed in disks \citep{ansdell16,Long17,miotello17,yu17}, (2) warmer temperatures (20 K $<$ T$_{\rm{gas}}$ $<$ 100 K) hinder CO processing \citep[e.g.,][]{reboussin15,schwarz18,Bosman18,schwarz19}, (3) higher cosmic ray rates lead to increased CO depletion \citep[e.g.,][]{reboussin15,Eistrup16,dodson18,schwarz18,eistrup18,Bosman18,schwarz19},  (4) CO depletion is enhanced over time \citep[e.g.,][]{reboussin15,schwarz18,zhang2020}.

ML techniques are thus a powerful complement to chemical modeling efforts. With simple algorithms like multiple linear and polynomial regression that fit our data within minutes, we have shown how ML can be exploited not only for predictions, but also for explanatory purposes. ML can handle complex multi-parameter models and the large amounts of resulting data in an efficient fashion. We believe with even more advanced techniques, we can further deepen our understanding of complex chemical processes.

It is important to note that the disk model used for this study presents some limitations. First of all, beyond the chemical evolution, our disk is static and does not account for changing physical conditions of the gas over time. Likewise, our model does not include dynamical processes such as radial drift, grain growth, and vertical mixing, which have been theoretically demonstrated to impact CO chemistry \citep{krijt18, krijt2020}. Nevertheless, this approach is flexible enough to include additional parameters for more complex models in the future. One possible expansion of this work is to include dust-related parameters such as dust sizes, compositions, temperatures, since the dust affects the disk opacity and is vital for grain-surface chemistry. In addition, this approach can be expanded to study disks with different physical structures as well as other molecules of interest. We work here with the main CO isotopologue, $^{12}$CO, but a similar approach can be taken for rarer isotopologues such as $^{13}$CO and C$^{18}$O.

\section{Conclusions} \label{sec:conclusion}
In this work, we test the ability of machine learning (ML) in helping us understand the interplay between chemistry and physical conditions in protoplanetary disks. Specifically, we investigate the impact of physical conditions, i.e., gas density, gas temperature, cosmic ray ionization rate, X-ray ionization rate, and UV flux on CO chemistry by tackling the diversity of chemical environments within a single disk using linear and polynomial regression. We also use the feature importance scores from random forests to understand the impact of the cosmic ray ionization rate, initial C/O, and disk age. We summarize our findings below:
\begin{itemize}
    \item Machine learning can be a highly efficient tool to understand the behavior of complex multi-parameter models, such as astrochemical simulations.
    \item A linear model does not accurately capture the underlying CO response to physical conditions in the disk model. We find that a polynomial model is required; however the number of terms makes such a model challenging to interpret and subject to multicollinearity. Using a combination of techniques, we are able to simplify our model, while maintaining its accuracy and making its meaning more transparent. The coefficient analysis reveals that there is a tendency toward CO depletion in the disk and that combination of physical parameters should be considered when conducting sensitivity analyses. Our analysis also reveals that the gas temperature and gas density are key factors that influence CO chemistry.
    \item CO chemistry in the warm molecular layer can be described with a multiple linear regression model, which suggests that all of the complex processes leading to CO depletion are reduced in this region.
    \item We note that an increased cosmic ray ionization rate and increased C/O ratio can lead to further depletion. The former effect does so by yielding more He$^+$ and H$_3^+$, which destroy CO and respectively lead to the formation of CH$_4$ and CH$_3$OH. The latter effect creates  an oxygen-poor environment that leads to carbon being locked up in hydrocarbons such as C$_2$H rather than CO.
    \item Over time, the terms involving the radiation field and cosmic rays become more significant, suggesting that CO is destroyed due to the build up species like He$^+$, H$_3^+$, OH that eventually contribute to the formation of CH$_4$, CH$_3$OH, CO$_2$.
\end{itemize}
Our ML-driven approach shows that CO chemistry is multifaceted, consistent with previous work. Thus, machine learning is a powerful tool that can complement our chemical modeling efforts by aiding the analysis of large data sets and exploration of large parameter spaces.

\begin{acknowledgments}
We thank the anonymous referees for providing helpful
comments. A.D. acknowledges support from the John. F. Angle Fellowship. L.I.C. acknowledges support from the David and Lucille Packard Foundation, the Research Corporation for Scientific Advancement Cottrell Scholar Award, and NASA ATP 80NSSC20K0529. J. Pegues is supported by an STScI Postdoctoral Fellowship. The National Radio Astronomy Observatory is a facility of the National Science Foundation operated under cooperative agreement by Associated Universities, Inc.
\end{acknowledgments}
\appendix
\twocolumngrid

As described in Section \ref{pol_reg}, the full polynomial model accurately captures the relationship between CO abundance and physical conditions in the disk environment. However, this model contains many terms, which makes interpretation challenging and it suffers from multicollinearity. Multicollinearity means that the independent variables/predictors are correlated with one another. Multicollinearity poses a challenge for our goal of building an explanatory model: it makes it difficult to isolate the impact of each predictor on CO without the influence of the remaining predictors. In this section, we explain how we reduce the number of terms in the polynomial model to facilitate interpretation and minimize multicollinearity. Specifically, we use random forests, centering, and a variation inflation factor analysis, which are discussed in more detail below. We ensure that the variation inflation factors for all terms in the ML model are below 10 to reduce the impact of multicollinearity (see Appendix B).

\medskip

\section{Decision trees and random forests}\label{rfs}
\begin{figure*}[htb!]
    \centering
    \includegraphics[width=\textwidth]{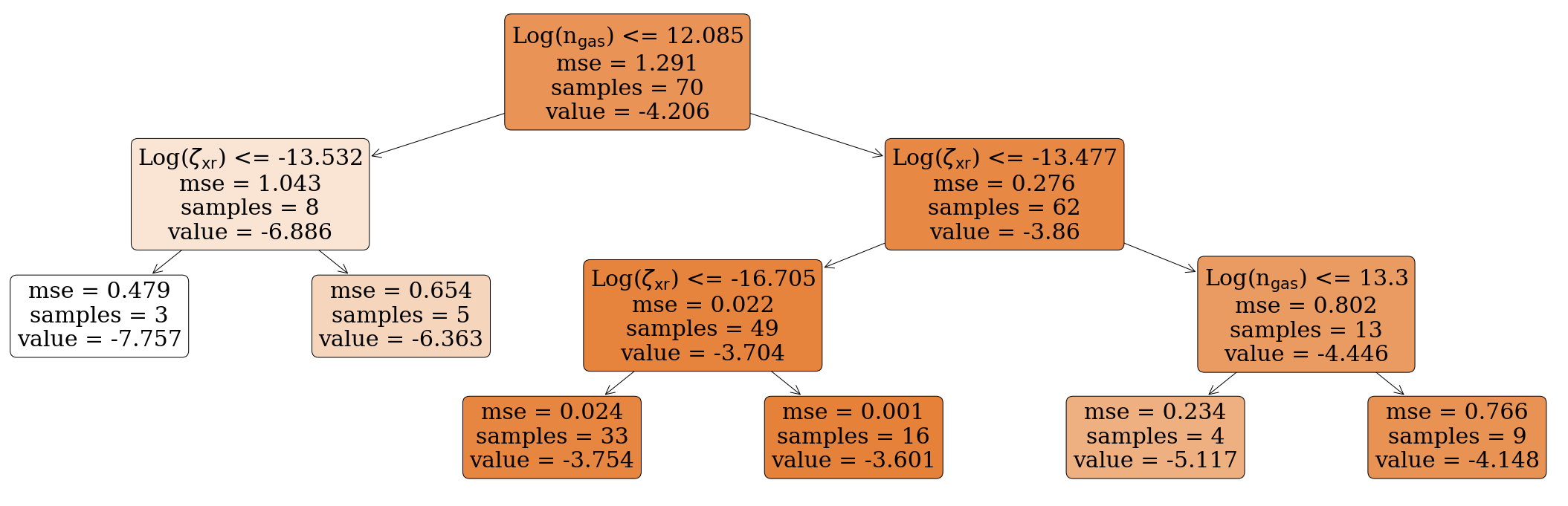}
    \caption{An example of a decision tree using 70 random points from our data. The top box represents the root node, the boxes in between the internal nodes, and the boxes  at the bottom the leaf nodes/leaves. Samples represent the number of points that belong to the node, value is the average of log(CO abundance) for points in the node, and mse is the mean square error, which the decision tree is trying to minimize. The values in the leaves represent the predictions. To determine the condition for the root node, the algorithm runs through all features and calculates possible thresholds to find the split that will minimize the mse. In this example, the splitting condition is log(n$_{\textrm{gas}}$) $<$ 12.085. Points that fulfill this condition go to the internal node on the left  and those that do not go to the node on the right: there are 8 samples that fulfill the condition and 62 that do not. The same process is applied to the internal nodes until no further split can be performed. To predict the value of a new point, the point is passed through the tree until it reaches the leaf it belongs to. }
    \label{fig:tree_toy}
\end{figure*}

A decision tree is a non-parametric ML algorithm and it owes its name to the fact that it has a tree-like structure. It is made up of nodes, which yield the tree-like structure. The top node where the first split is performed is called the root node, the nodes within the tree are the internal nodes, and the terminal nodes where predictions are made are the leaf nodes or leaves. To determine the condition for the root node, the algorithm runs through all features and calculates possible thresholds to find the split that will minimize the mean squared error (mse). The best feature and threshold are then used for the first splitting condition, i.e., \emph{feature X} $<$ \emph{threshold}. The split data set is  then iteratively divided following a similar procedure until no further split can be performed. The prediction of a leaf is the average of the data points in that leaf. To predict the value of a new point, the point is passed through the tree until it reaches the leaf where it belongs to. Figure \ref{fig:tree_toy} shows an example of a decision tree.
Decision trees are prone to overfitting thus Random forest \citep{ho95, Ho98, Breiman01} (RFs) are used since they overcome this issue by averaging the predictions of multiple trees.

An useful feature of RFs is the feature importance score, which is a dimensionless number between 0 and 1 that expresses how efficiently each predictor splits parameter space in the RF. The higher the score, the more important the predictor is. The scores of all predictors are normalized to sum up to 1. We use this to determine the most important terms in our polynomial model to minimize multicollinearity as well as to compare the fiducial disk model with the high cosmic ray and high C/O models and the models at different ages.

\section{Tackling multicollinearity}\label{mult}
We search for multicollinearity by looking at the variation inflation factor (VIF) of each predictor. The VIF is a dimensionless metric expressed as:
\begin{equation}
    \textrm{VIF} = \frac{1}{1-\textrm{R}_i^2}
\end{equation}
where R$_i^2$ is the R$^2$ score obtained when the i$^{\text{th}}$ predictor is regressed on the other predictors. The higher the VIF of a predictor, the more that predictor is correlated with other predictors. As a rule of thumb, a VIF of 1 indicates no correlation, a VIF between 1 and 5 indicates low correlation, a VIF between 5 and 10 indicates moderate correlation, and a VIF greater than 10 indicates serious multicollinearity.  Our initial predictors (T$_{\textrm{gas}}$, n$_{\textrm{gas}}$, $\zeta_{\textrm{xr}}$, $\zeta_{\textrm{cr}}$, and F$_{\textrm{uv}}$) display negligible multicollinearity since their VIFs are less than 5 (Table \ref{tab:vif}).

\begin{deluxetable}{cc}[htb!]
\tablecaption{Initial predictors and their corresponding variation inflation factors. \label{tab:vif}}
\tablehead{\colhead{Predictor} & \colhead{Variation inflation factor}}
\decimals
\startdata
Log$_{10}$(n$_{\textrm{gas}}$) & 3.82 \\
Log$_{10}$(T$_{\textrm{gas}}$) & 2.23 \\
Log$_{10}$(F$_{\textrm{uv}}$) & 3.00 \\
Log$_{10}$($\zeta_{\textrm{cr}}$) & 3.43\\
Log$_{10}$($\zeta_{\textrm{xr}}$) & 2.67\\
\enddata

\end{deluxetable}
\begin{figure}[h!]
    \centering
    \includegraphics[scale=0.3]{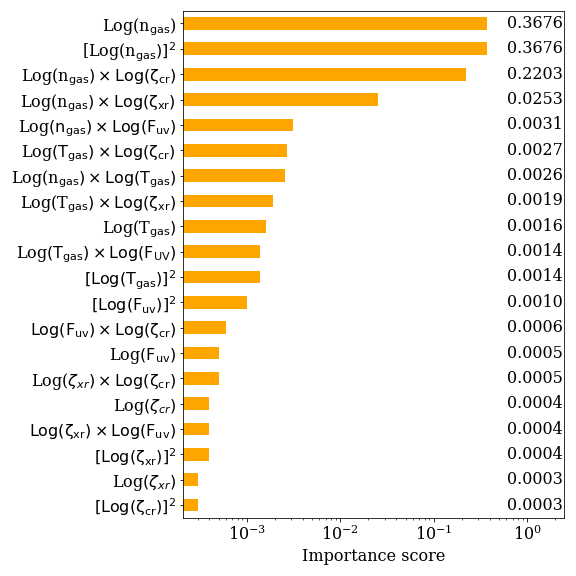}
    \caption{Importance scores for the full polynomial model.}
    \label{fig:feature_importance}
\end{figure}

However, once the predictors undergo the polynomial transformation, structural multicollinearity arises (Table \ref{tab:vif_cent_uncent}). This multicollinearity is due to the creation of new predictors using existing predictors, e.g., generating X$^2$ using X, generating the cross-term X$_1$X$_2$ using X$_1$ and X$_2$. Multicollinearity is not an issue when the goal is to make predictions. However, since we want to isolate the contribution of each predictor, we need to minimize multicollinearity as much as possible. We first reduce it by centering the initial predictors (T$_{\textrm{gas}}$, n$_{\textrm{gas}}$, $\zeta_{\textrm{xr}}$, $\zeta_{\textrm{cr}}$, and F$_{\textrm{uv}}$). Centering is a technique that consists in taking the original predictors, calculating their respective means, and subtracting those mean values from each data point. The polynomial transformation is then applied to the centered data points. Mathematically,  centering is expressed as:
\begin{equation}
    \textrm{X}_{\textrm{cent},i} = \textrm{X}_i - \textrm{mean(X)}
\end{equation}
Centered terms in this work are denoted as $\overline{\textrm{X}}$. For example:
\begin{equation}
    \overline{\textrm{log}(\textrm{T}_{gas})} = \textrm{log}(\textrm{T}_{gas}) - \textrm{mean}[\textrm{log}(\textrm{T}_{gas})]
\end{equation}

Centering drastically reduces multicollinearity as noted in Table \ref{tab:vif_cent_uncent}. To further reduce it, we exploit the feature importance score from random forests to find the most important terms in the polynomial. The resulting scores are shown in Figure \ref{fig:feature_importance}. We then incorporate terms one by one in the polynomial using their importance scores--following a decreasing order--until we reach the point where the VIF of one of the predictors exceeds 10, indicating the presence of multicollinearity. We then only keep the terms from the previous step. The VIF of the predictors in the simplified polynomial are shown in Table \ref{tab:vif_simp_pol}, which indicates that multicollinearity is minimized.

\begin{deluxetable*}{cccc}[htb!]
\tablecaption{Variation inflation factors for the raw and centered predictors in the polynomial model. $\overline{\textrm{X}}$ denotes the centered predictors. T1-T8 denote the terms in the simplified polynomial model.
\label{tab:vif_cent_uncent}}
\tablehead{\colhead{Predictor} & \colhead{Variation inflation factor} & \colhead{Predictor} &  \colhead{Variation inflation factor}}
\decimals
\startdata
Log (n${\textrm{gas}}$) & 500376 & $\overline{\textrm{Log} (\textrm{n}{\textrm{gas}})}$ &  158\\
Log ($\zeta_{\textrm{xr}}$) & 204868 & $\overline{\textrm{Log} (\zeta_{\textrm{xr}}}$) & 112\\
Log (F${\textrm{uv}}$) & 549602 &  $\overline{\textrm{Log} (\textrm{F}{\textrm{uv}})}$ & 237 \\
Log ($\zeta_{\textrm{cr}}$) & 2018 & $\overline{\textrm{Log} (\zeta_{\textrm{cr}}}$) & 104\\
$\lbrack$ Log$(\textrm{n}_{\textrm{gas}}$)$\rbrack^2$ & 2495 & $\overline{\lbrack \textrm{Log}
(\textrm{n}{\textrm{gas}})\rbrack^2}$ & 31 \\
$\lbrack$ Log (T${\textrm{gas}}$)\rbrack$^2$ & 586  & $\overline{\lbrack \textrm{Log} (\textrm{T}{\textrm{gas}})\rbrack^2}$ & 16  \\
$\lbrack$ Log ($\zeta{\textrm{xr}}$)\rbrack$^2$ & 1510 & $\overline{\lbrack \textrm{Log} (\zeta_{\textrm{xr}})\rbrack^2}$ & 71 \\
$\lbrack$ Log (F${\textrm{uv}}$)\rbrack$^2$ & 65 & $\overline{\lbrack \textrm{Log} (\textrm{F}{\textrm{uv}})\rbrack^2}$ & 29 \\
$\lbrack$ Log ($\zeta_{\textrm{cr}}$)\rbrack$^2$ & 3118 & $\overline{\lbrack \textrm{Log} (\zeta_{\textrm{cr}})\rbrack^2}$  & 25 \\
Log ($\zeta_{\textrm{xr}}$) $\times$ Log (F${\textrm{uv}}$) & 2255 & $\overline{\textrm{Log} (\zeta{\textrm{xr}})}$ $\times$ $\overline{\textrm{Log} (\textrm{F}{\textrm{uv}})}$ & 123\\
Log (F${\textrm{uv}}$) $\times$ Log ($\zeta_{\textrm{cr}}$) & 569802 & $\overline{\textrm{Log} (\textrm{F}{\textrm{uv}})}$ $\times$ $\overline{\textrm{Log} (\zeta{\textrm{cr}})}$ & 36 \\
Log ($\zeta_{\textrm{xr}}$) $\times$ Log ($\zeta_{\textrm{cr}}$) & 247341 & $\overline{\textrm{Log} (\zeta_{\textrm{xr}})}$ $\times$ $\overline{\textrm{Log} (\zeta_{\textrm{cr}})}$ & 49 \\
Log (n${\textrm{gas}}$) $\times$ Log (T${\textrm{gas}}$) [T1]  & 741 & $\overline{\textrm{Log} (\textrm{n}{\textrm{gas}})}$ $\times$ $\overline{\textrm{Log} (\textrm{T}{\textrm{gas}})}$  & 29\\
Log (T${\textrm{gas}}$) $\times$ Log (F${\textrm{uv}}$) [T2]  & 4647 & $\overline{\textrm{Log} (\textrm{T}{\textrm{gas}})}$ $\times$ $\overline{\textrm{Log} (\textrm{F}{\textrm{uv}})}$  & 192\\
Log (T${\textrm{gas}}$) $\times$ Log ($\zeta{\textrm{cr}}$) [T3] & 28125 & $\overline{\textrm{Log} (\textrm{T}{\textrm{gas}})}$ $\times$ $\overline{\textrm{Log} (\zeta{\textrm{cr}})}$   & 19\\
Log (n${\textrm{gas}}$) $\times$ Log ($\zeta{\textrm{xr}}$) [T4] & 8840  & $\overline{\textrm{Log} (\textrm{n}{\textrm{gas}})}$ $\times$ $\overline{\textrm{Log} (\zeta{\textrm{xr}})}$  & 158\\
Log (T${\textrm{gas}}$) [T5] & 24205 & $\overline{\textrm{Log} (\textrm{T}{\textrm{gas}})}$ & 107\\
Log (n${\textrm{gas}}$) $\times$ Log ($\zeta{\textrm{cr}}$) [T6]  & 534369   & $\overline{\textrm{Log} (\textrm{n}{\textrm{gas}})}$ $\times$ $\overline{\textrm{Log} (\zeta{\textrm{cr}})}$  & 104\\
Log (n${\textrm{gas}}$) $\times$ Log (F${\textrm{uv}}$) [T7]  & 11631 & $\overline{\textrm{Log} (\textrm{n}{\textrm{gas}})}$ $\times$ $\overline{\textrm{Log} (\textrm{F}{\textrm{uv}})}$ & 185\\
Log (T${\textrm{gas}}$) $\times$ Log ($\zeta{\textrm{xr}}$) [T8]  & 1281 & $\overline{\textrm{Log} (\textrm{T}{\textrm{gas}})}$ $\times$ $\overline{\textrm{Log} (\zeta{\textrm{xr}})}$  & 62\\
\enddata
\end{deluxetable*}

\begin{deluxetable}{cc}[htb!]
\tablecaption{Variation inflation factors for the predictors in the simplified polynomial. T1-T8 denote the terms in the simplified polynomial model.
\label{tab:vif_simp_pol}}
\tablehead{\colhead{Predictor} & \colhead{Variation inflation factor}}
\decimals
\startdata
$\overline{\textrm{Log} (\textrm{n}{\textrm{gas}})}$ &  2.56\\
$\overline{\lbrack \textrm{Log} (\textrm{n}{\textrm{gas}})\rbrack^2}$ & 7.90 \\
$\overline{\textrm{Log} (\textrm{n}{\textrm{gas}})}$ $\times$ $\overline{\textrm{Log} (\textrm{T}{\textrm{gas}})}$ [T1] & 7.80\\
$\overline{\textrm{Log} (\textrm{T}{\textrm{gas}})}$ $\times$ $\overline{\textrm{Log} (\textrm{F}{\textrm{uv}})}$ [T2]  & 4.14\\
$\overline{\textrm{Log} (\textrm{T}{\textrm{gas}})}$ $\times$ $\overline{\textrm{Log} (\zeta{\textrm{cr}})}$ [T3]   & 4.63\\
$\overline{\textrm{Log} (\textrm{n}{\textrm{gas}})}$ $\times$ $\overline{\textrm{Log} (\zeta{\textrm{xr}})}$ [T4]  & 3.27\\
$\overline{\textrm{Log} (\textrm{T}{\textrm{gas}})}$ [T5]  & 2.75\\
$\overline{\textrm{Log} (\textrm{n}{\textrm{gas}})} \times \overline{\textrm{Log} (\zeta_{\textrm{cr}})}$ [T6] & 6.30\\
$\overline{\textrm{Log} (\textrm{n}{\textrm{gas}})}$ $\times$ $\overline{\textrm{Log} (\textrm{F}{\textrm{uv}})}$ [T7] & 5.47\\
$\overline{\textrm{Log} (\textrm{T}{\textrm{gas}})}$ $\times$ $\overline{\textrm{Log} (\zeta{\textrm{xr}})}$ [T8]  & 5.62\\
\enddata
\end{deluxetable}

\end{document}